\documentclass[journal=jacsat,manuscript=article]{achemso}

\usepackage[version=3]{mhchem} 
\usepackage{todonotes}
\usepackage{booktabs}
\usepackage{multirow}
\usepackage{graphicx}
\usepackage{caption}
\usepackage{subcaption}
\usepackage{xcolor}
\usepackage{url}

\usepackage{xspace}

\author{Yachan Liu}
\affiliation[University of Massachusetts Amherst]
{Department of Chemical and Biomolecular Engineering, University of Massachusetts Amherst, Amherst, Massachusetts 01003, United States of America}
\author{Elaine Wu}
\affiliation[University of Massachusetts Amherst]
{Department of Chemical and Biomolecular Engineering, University of Massachusetts Amherst, Amherst, Massachusetts 01003, United States of America}
\author{Ping Yang}
\affiliation[University of Massachusetts Amherst]
{Department of Chemical and Biomolecular Engineering, University of Massachusetts Amherst, Amherst, Massachusetts 01003, United States of America}
\author{Aaron Sun}
\affiliation[University of Massachusetts Amherst]
{College of Information and Computer Sciences, University of Massachusetts Amherst, Amherst, Massachusetts 01003, United States of America}
\author{Subhransu Maji}
\affiliation[University of Massachusetts Amherst]
{College of Information and Computer Sciences, University of Massachusetts Amherst, Amherst, Massachusetts 01003, United States of America}
\author{Wei Fan}
\affiliation[University of Massachusetts Amherst]
{Department of Chemical and Biomolecular Engineering, University of Massachusetts Amherst, Amherst, Massachusetts 01003, United States of America}
\author{Peng Bai}
\email{pengbai@umass.edu}
\phone{+1 413 545 6189}
\fax{+1 413 545 1647}
\affiliation[University of Massachusetts Amherst]
{Department of Chemical and Biomolecular Engineering, University of Massachusetts Amherst, Amherst, Massachusetts 01003, United States of America}

\title[An \textsf{achemso} demo]
  {Assessment of the synthetic feasibility of hypothetical zeolite-like materials based on ZeoNet}

\abbreviations{IR,NMR,UV}

\begin{document}

\newpage
\begin{abstract}
A suite of classifiers was developed to distinguish experimentally synthesized zeolites from computationally predicted zeolite-like structures. Using convolutional neural networks applied to 3D volumetric grids, these classifiers achieve accuracies more than an order of magnitude higher than previous approaches based on geometric filters or other machine learning methods. The best-performing model differentiates among hypothetical zeolites and those that can be synthesized as silicates, as aluminophosphates, or as both. This four-class classifier attains a false negative rate of 3.4\% and a false positive rate of 0.4\%, misidentifying only 1,207 of over 330,000 hypothetical structures—even though the hypothetical structures exhibit similar formation energies as real zeolites and chemically reasonable bond lengths and angles. We hypothesize that the ZeoNet representation captures essential structural features correlated with synthetic feasibility. In the absence of comprehensive physics-based criteria for synthesizability, the small subset of misclassified hypothetical structures likely represents promising candidates for future experimental synthesis.
\end{abstract}

\section{TOC Graphic}
\begin{center}
    \includegraphics[width=5.5cm]{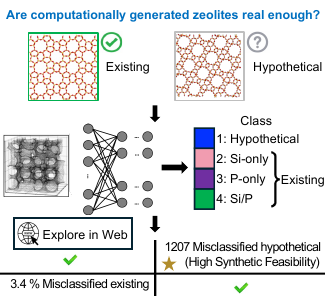}
\end{center}
\newpage

Zeolites are silicates or aluminophosphates consisting of corner-sharing TO$_4$ tetrahedra (where T = Si, Al, or P) that organize into a crystalline porous framework. The molecularly sized internal cavity allows these materials to function as shape-selective catalysts, sorbents, and ion-exchangers.\cite{maesen2007zeolite} Apart from their established roles in the petroleum industry, zeolites are also being investigated for a broad range of emerging applications for renewable energy and sustainability, such as membrane separations, biomass conversion, and plastic waste upcycling. To date, 260 known framework topologies have been catalogued by the International Zeolite Association (IZA),\cite{IZA-SC} yet the space of potential zeolites is, in principle, infinite, with $10^5$--$10^6$ structures enumerated computationally.\cite{falcioni1999biased, treacy2004enumeration, foster2004chemically, woodley2004prediction, pophale2011database, li2012fragen, abdelkafi2017using, liu2017determining} Large-scale, in-silico screening has further revealed tantalizing performance of hypothetical structures that drastically exceeds that of existing zeolites.\cite{haldoupis2011pore, lin2012silico, kim2012large,  kim2013large, bai2015discovery} However, it is nearly impossible at present to realize new zeolite structures through rational synthesis planning.\cite{cundy2005hydrothermal, coronas2010present} Limited rules of thumb exist, but synthesizing a new structure, or even broadening the synthesis scope of an existing structure, requires significant trial-and-error. As a result, a handful of new structures are discovered and added to the IZA database each year. It is unclear whether most hypothetical zeolites do not yet exist simply because their synthesis recipes have not been found or because there are intrinsic, hitherto unknown reasons that prevent their synthesis, however hard one may try.

To address the question of synthesizability, several approaches have been attempted. To begin, the construction of hypothetical structure databases already included an energy criterion to remove structures with formation energies that are too high. The threshold values were usually chosen based on the formation energies of existing zeolites (e.g., 30~kJ/mol Si relative to $\alpha$-quartz, the most stable polymorph of tectosilicates; see Figure~S1).\cite{foster2004chemically, pophale2011database} Later studies analyzed geometric parameters such as framework density, the distance between a tetrahedral site (T-site; e.g., as occupied by Si or Al) and its 5th neighbor, and TTT angles. By considering the distribution of these geometric parameters for existing zeolites, up to 50\% of hypothetical structures in the Predicted Crystallography Open Database (PCOD) were suggested to be unfeasible.\cite{li2013criteria, salcedo2019high, li2019necessity} Finally, more recent efforts applied data science methods to assess structural similarity.\cite{carr2009machine, muraoka2019linking, ma2020thermodynamic, helfrecht2022ranking} For example, Muraoka et al.\cite{muraoka2019linking} extracted synthetic descriptors from the literature using machine learning and converted them to structural descriptors that indicate the presence or absence of building units. These structural descriptors were then used to construct a similarity map of crystal structures. Helfrecht et al.\cite{helfrecht2022ranking} used the Smooth Overlap of Atomic Positions (SOAP) method to describe the local environment of framework atoms and structural correlations. A linear Support Vector Machine (SVM) classifier was then trained to distinguish between real and hypothetical zeolites, which achieved an accuracy of 89\% for IZA and 95\% for PCOD structures.

\begin{figure*}
    \includegraphics[width=17.4cm]{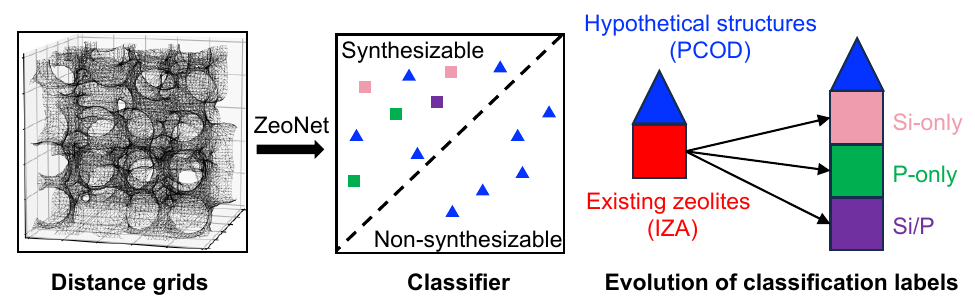}
    \caption{The ZeoNet pipeline for predicting synthetic feasibility in zeolites.}
    \label{fig:model_diagram}
\end{figure*}

In this work, we trained a series of classifiers based on ZeoNet,\cite{liu2023zeonet} a recently developed 3-dimensional convolutional neural network (ConvNet) representation, to predict the synthesizability of a given framework structure (Figure~\ref{fig:model_diagram}). Our dataset consists of hypothetical structures from the PCOD database and existing zeolites from the IZA database. To construct the input representation for ZeoNet, a volumetric distance grid is calculated by setting the radii of tetrahedral atoms and bridging atoms to those of silicon and oxygen, respectively, effectively treating all IZA and PCOD structures as siliceous. This choice was made for consistency with the hydrocarbon adsorption tasks used for pretraining ZeoNet.\cite{bai2015discovery} The classifier is forced to predict all PCOD entries as unsynthesizable and IZA entries as synthesizable. Clearly, such a classifier cannot be perfect, as the PCOD database contains structures that have been or may eventually be synthesized. Our hypothesis is that the misclassified hypothetical structures have higher synthetic feasibility.

To test this hypothesis, we started by fine-tuning ZeoNet trained for predicting long-chain hydrocarbon adsorption.\cite{liu2023zeonet,liu2025representation} A new fully connected (FC) layer with two neurons was randomly initialized to replace the final layer and trained from scratch using a learning rate of 0.01, while all other layers and their weights were retained and updated using a smaller learning rate of 0.001. The raw scores from the FC layer were passed through a softmax activation function to generate a probability distribution over the target classes. The class with the highest predicted probability was selected as the classifier output. The success of this approach depends on the \textit{assumption} that the structural features optimized by ZeoNet for adsorption performance correlate with thermodynamic or kinetic feasibility of the material. This is a much more challenging cross-domain transfer learning,\cite{gupta2021cross, kong2021materials, chen2024bulk} compared to transfer learning to different adsorption tasks.\cite{liu2025representation}

\begin{figure}[ht]
    \includegraphics[width=6.5cm]{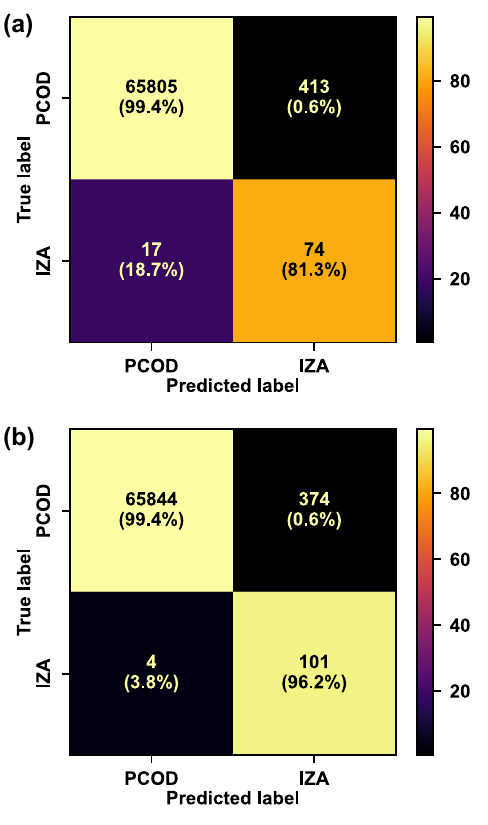}
    \caption{Confusion matrix for (a) the two-class model and (b) phosphate-aware two-class model evaluated on the validation set. Percentages are indicated by color and shown in parentheses.}
    \label{fig:2class}
\end{figure}

Figure~\ref{fig:2class}a shows the confusion matrix for the resulting binary classifier evaluated on the validation set. Surprisingly, the classifier misidentifies only 0.6\% of PCOD structures, which is substantially better than that achieved with geometric filters\cite{li2013criteria, salcedo2019high, li2019necessity} and an order of magnitude smaller than even the SOAP-based SVM classifier.\cite{helfrecht2022ranking} On the other hand, the model only shows an accuracy of 81.3\% for the IZA structures, incorrectly labeling a number of existing zeolites as unsynthesizable. A detailed examination of the 17 false negative IZA structures in the validation set revealed that nearly 50\% had never been synthesized as (alumino)silicates, compared to about 20\% in the whole IZA database. Furthermore, it is particularly interesting to note that at least two of them, IM-6, a cobalt–gallium phosphate with the framework code USI, and SU-16, a borogermanate with the framework code SOS, are both unstable upon the removal of organic structure-directing agents.\cite{josien2003synthesis, li200516}

These observations bolster our confidence that the ZeoNet representation may have captured essential structural characteristics of zeolites that correlate with synthetic feasibility. To distinguish silicates from aluminophosphates, the second largest category of materials in the IZA database, we added a single binary input variable, $c$, to indicate the nature of the framework chemistry (see Table~S1). The ConvNet feature vector is concatenated with $c$ before passing through the final FC layer. If an IZA structure has been synthesized in both silicate and aluminophosphate forms, the structure will appear twice in the training set, each with their respective values of $c$. For PCOD structures, $c$ is randomly assigned one of the two values to avoid further increasing the number of negative samples in the training set. As shown in Figure~\ref{fig:2class}b, the additional indicator significantly improves the classifier's sensitivity, with the number of false negative instances reduced from 18.7\% to 3.8\%, incorrectly labeling only four out of 105 IZA structures, while the specificity of the model remains unchanged for the PCOD structures. This finding strongly suggests that silicates and aluminophosphates are structurally distinct and that it is essential for a classifier to distinguish between the two forms. By combining the distance grid for a hypothetical structure with the additional indicator, the model can further predict the likelihood of synthesizing that structure in both forms. For the $c$ assignment used in Figure~\ref{fig:2class}b, 199 of the 374 false positives are silicates while 175 are aluminophosphates.

\begin{figure}[htb]
    \includegraphics[width=6.5cm]{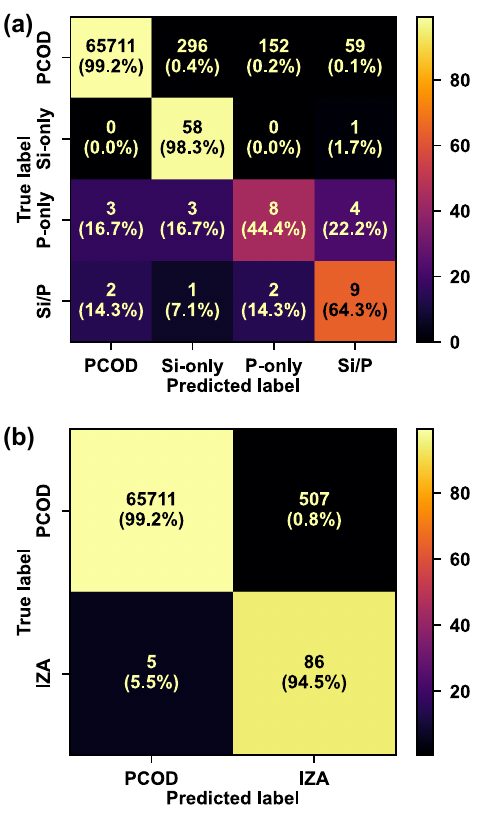}
    \caption{Validation performance of the four-class model, showing its (a) confusion matrix and (b) effective accuracy on predicting an overall synthetic feasibility by merging Si-only, P-only, and Si/P classes into a single IZA class. Percentages are indicated by color and shown in parentheses.}
    \label{fig:4class}
\end{figure}

Despite the increased overall accuracy of the composition-aware classifier, the lumped class labels do not allow for the assessment of the classifier performance for silicates and aluminophosphates separately. To address this issue, we tested a multi-class model by training it to sort IZA structures into synthesizable as (alumino)silicates (labeled as Si-only), as (silico)aluminophosphates (labeled as P-only), or as both forms (labeled as Si/P), while still predicting all PCOD structures as unsynthesizable. This model does not require the assignment of silicates or aluminophosphates to the PCOD structures during training but predict this label in the output class. The resulting confusion matrix is given in Figure~\ref{fig:4class}a. To compare with the binary classifiers, Figure~\ref{fig:4class}b shows an effective two-class confusion matrix by merging the three synthesizable classes. The four-class model performs similarly as the composition-aware binary model, with only slightly higher false positive and false negative rates. The breakdown of the classifier performance shows a very high accuracy (98.3\%) for Si-only structures, mistaking only 1.7\% as synthesizable as dual Si/P forms. The P-only structures, in contrast, exhibit a much lower accuracy (44.4\%), with 22.2\% of Pi-only structures predicted to be synthesizable as dual Si/P forms and the rest predicted to be either Si-only or unsynthesizable. Nonetheless, despite the mistaken detailed labeling, the four-class classifier does still correctly identify 83.3\% of the P-only structures to be synthesizable.

\begin{figure}[t]
    \includegraphics[width=6.5cm]{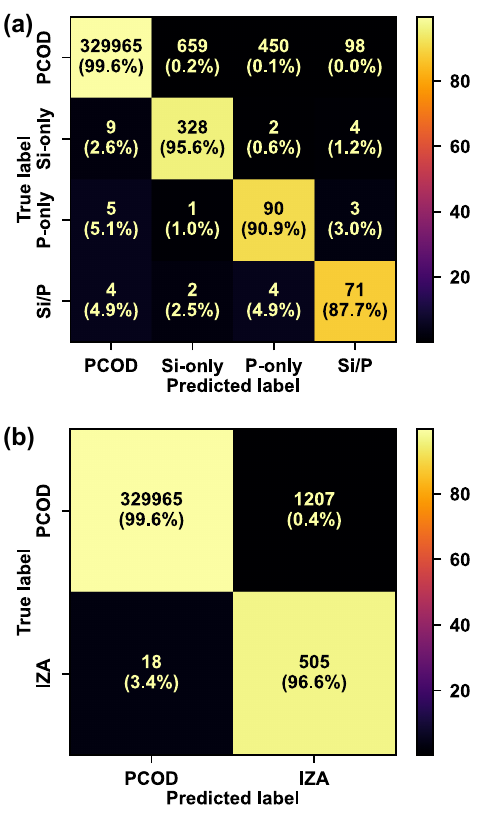}
    \caption{Performance of the four-class model on the full dataset: (a) confusion matrix and (b) effective accuracy on predicting an overall synthetic feasibility by merging Si-only, P-only, and Si/P classes into a single IZA class. Percentages are indicated by color and shown in parentheses.}
    \label{fig:4class-extended}
\end{figure}

The poor performance likely reflects the under-representation of aluminophosphates in the IZA database. This is especially true for the version used to train the original ZeoNet, which includes 383 materials across 204 framework types (not including structures that do not have the SiO$_2$ stoichiometry). An updated version as of April 2025 increases to 523 structures across 246 frameworks. We retrained the four-class model on the updated IZA database and evaluated the best model on the full dataset. As shown in Figure~\ref{fig:4class-extended}, the final model achieved excellent performance for all four class labels, with significantly improved accuracy, 90.9\% and 87.7\% respectively, for P-only and Si/P structures. Since the classifier architecture remains unchanged, this performance improvement reflects solely the impact of additional data in the low-data regime. All misclassified IZA structures are listed in Figure~S2. A close examination of these structures reveals potentially useful information. For example, Si-only and P-only structures that are misidentified as dual Si/P can be a "temporary error", as it is conceivable that synthetic advancement may allow these zeolites to be synthesized in the other form, thus becoming dual Si/P. The overall accuracy for the merged IZA category is 96.6\%, while only 0.4\% of PCOD structures (1207 out of 331,172; Table~S2) were falsely identified to be IZA-like. We speculate that these misclassified PCOD structures are prime candidates as future synthetic targets. They are listed in Table~S3, sorted by their model-predicted class probability.

\begin{figure*}[htb]
    \includegraphics[width=15cm]{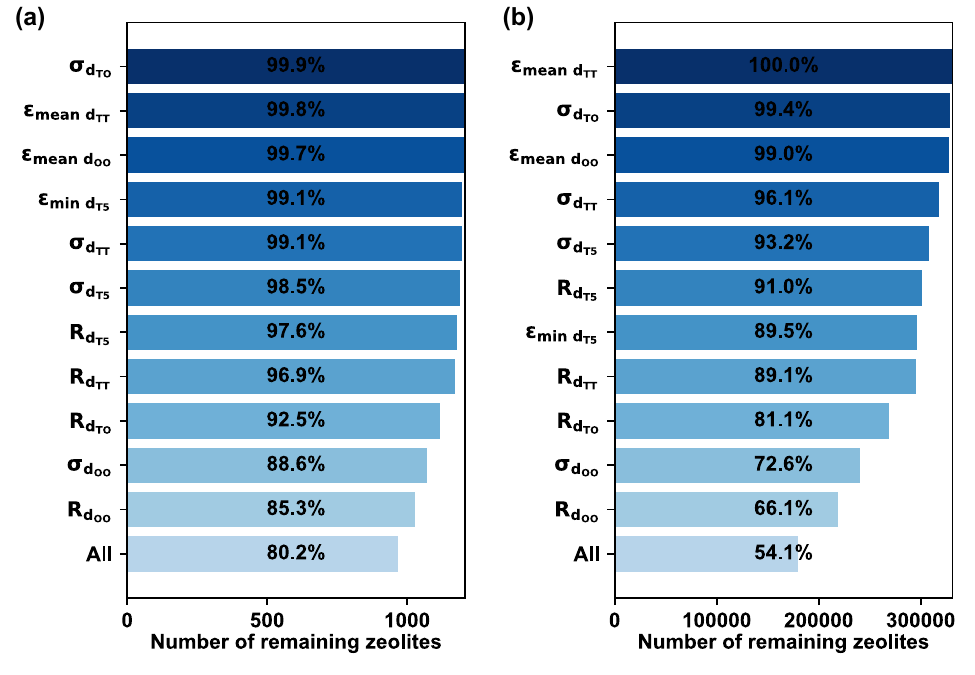}
    \caption{The number and percentage of zeolites passing each individual filter for (a) misclassified PCOD structures and (b) all PCOD structures. Each row shows the residual rate of applying a single filter criterion while the combined residual rate is given in the last row.}
    \label{fig:filter}
\end{figure*}

It is informative to examine these misclassified structures using geometric descriptors previously proposed for filtering hypothetical structures. Here, we explored the approach of Perez et al.,\cite{salcedo2019high} which utilizes four types of local interatomic distances (LIDs) including T–O distances ($d_\mathrm{TO}$), T–(O)–T distances ($d_\mathrm{TT}$), O–(T)–O distances ($d_\mathrm{OO}$), and the distances to the fifth nearest neighbor of T atoms ($d_\mathrm{T5}$; which should be a T atom bonded to one of the four bridging O atoms). Figure~S3 compares the distribution of these LIDs for the IZA and PCOD databases, which clearly shows that these two databases feature different geometric compositions. On average, the PCOD database contains more structures with longer TO and OO distances, but shorter TT and T5 distances. A PCOD structure is also more likely to exhibit a wider range of values for all four LIDs, as reflected by the larger standard deviations and ranges. These observations form the basis for geometric filters: the characteristics of IZA zeolites establish what's "normal" and a hypothetical structure is considered unfeasible if it significantly deviates from the IZA normality. As described in Methods, Perez et al.,\cite{salcedo2019high} assembled the four LIDs into 11 criteria, which can be grouped under three categories. Categories II and III (denoted as $\sigma_{d_\mathrm{AB}}$ and $\mathrm{R}_{d_\mathrm{AB}}$) simply demand that the standard deviations and ranges of the four LIDs for a candidate structure must not exceed the largest standard deviation and range that have ever been observed for a real (IZA) zeolite. The candidate is discarded if it fails any of the eight rules. In addition, as shown in Figure~S4, three pairs of LIDs were found to be highly correlated for IZA zeolites, while they are much more scattered for PCOD structures. These correlations were fitted via linear regressions and the largest deviations from the linear fits by any IZA zeolite are used as the upper bound that a feasible hypothetical structure must not exceed (Category I criteria, denoted as $\epsilon_{\mathrm{mean/min }\ d_\mathrm{AB}}$). Figure~\ref{fig:filter} shows the result of applying the 11 filters. Despite the apparent large differences in the distributions, the correlation between mean TT and TO distances is almost non-discriminative, which retains nearly 100\% of all PCOD structures. This reflects the outsized impact a few existing zeolites with rare geometric characteristics (see Figure~S4b) can have on a filter criterion. Similar observations can be made for the standard deviation of TO distances (Figure~S3a) and the correlation between mean OO and TO distances (Figure~S4a), both of which remove less than 1\% of PCOD structures. Interestingly, in contrast to $\sigma_{d_\mathrm{TO}}$, the range of TO distances in a given structure ($\mathrm{R}_{d_\mathrm{TO}}$) filters out 19\% of PCOD structures. Both standard deviation and range characterize the scatter; this contrast therefore highlights the sensitivity of geometric filters on the idiosyncrasy of the (evolving) IZA database. The most discriminative filter criteria are the standard deviation and range of OO distances, both of which remove around 30\% of PCOD structures and 11 -- 15\% of misclassified PCOD structures. In total, 179,230 out of 331,172 remain for the full PCOD database, while 968 out of the 1207 misclassified PCOD structures pass all geometric filters. The much higher combined success rate for the misclassified PCOD structures (80.2\% vs. 54.1\%) reflect the closer resemblance of these structures to the IZA zeolites than the PCOD database as a whole (Figures~S3--4). However, even the misclassified PCOD structures still exhibit geometric characteristics that push the boundary of known zeolites. Similar observations can be made for other geometric features (see Figure~S6): While the full PCOD database contains many more structures with smaller pores, higher framework atom density, lower surface area, and smaller pore volume, the misclassified PCOD structures exhibit much similar distributions as the IZA database. Finally, reducing the dimensionality of the ZeoNet features using techniques such as t-SNE\cite{JMLR:v9:vandermaaten08a} does not appear to yield useful low-dimensional partitions (Figure~S5).

In summary, a series of classifiers based on ZeoNets, which are ConvNets using volumetric distance grids and pretrained on the adsorption of long-chain hydrocarbon molecules,\cite{liu2023zeonet,liu2025representation} were developed to distinguish between real zeolites catalogued in the IZA database and hypothetical zeolite-like materials in the PCOD database. These classifiers achieve more than an order-of-magnitude improvement in accuracy over the best previous results\cite{li2013criteria, salcedo2019high, li2019necessity,carr2009machine, muraoka2019linking, ma2020thermodynamic, helfrecht2022ranking}. By separating (silico) aluminophosph\-ates from (alumino)silicates, the four-class classifier reaches a false negative rate of 3.4\% and a false positive rate of 0.4\%, misidentifying only 1207 PCOD structures as feasible. While geometric filters remain useful criteria that are physically meaningful, feature extraction by ZeoNet appears less sensitive to the changing characteristics of existing zeolites, as new materials are continuously being discovered, and is likely more capable of considering nonlinear, high-dimensional feature combinations. It is worth emphasizing, however, that the model does not predict thermodynamic or kinetic feasibility, but infers synthesizability based on the similarity with existing materials. In the absence of comprehensive physics-based understanding of synthesizability, we cautiously speculate that these misclassified hypothetical structures are the most promising candidates for new zeolite synthesis efforts. Towards this end, a web application is provided as Supporting Information that allows the exploration of our results in application-specific contexts where computational screening has identified high-performing hypothetical candidate structures.\cite{bai2015discovery, hewitt2022machine, kim2013new, kristof2024assessment, simon2015best} 

\section{Computational Methods}
\subsection{Zeolite dataset}
Computationally predicted zeolite-like materials were taken from the PCOD database,\cite{pophale2011database} while real zeolites were collected from the IZA website\cite{IZA-SC}, including both idealized framework structures and experimentally determined material structures. Structure that do not have the SiO$_2$ stoichiometry are excluded. The original ZeoNet and the various classifiers were initially trained on the 2014 version of the IZA database and the four-class classifier were refined on the 2025 version of the database. The structures were randomly split into training, validation, and test sets in a 7/2/1 ratio to monitor and control for overfitting. Since some IZA framework codes appear in multiple datasets, a likely better alternative would be to split the dataset based on framework codes, but this choice is unlikely to affect the false positive rates (i.e., misclassified PCOD structures). The IZA framework types (as indicated by a three-letter code) were sub-classified into synthesizable as (alumino)silicates (labeled as Si-only), as (silico)aluminophosphates (labeled as P-only), or as both forms (labeled as Si/P), based on the list of references given for each framework type on the IZA website. The same classification is given to all type materials under a given three-letter code; e.g., Si/P was assigned to both SSZ-13 and AlPO-34 under the framework code CHA. It is also possible to assign separate labels to the different structures belonging to the same framework code. A detailed breakdown is given in Table~\ref{tab:class_count}.

\begin{table}[]
\centering
\caption{Classification of IZA framework types and structures based on reported syntheses.}
\label{tab:class_count}
\begin{tabular}{c|c|c|c|c}
\hline
\multicolumn{5}{l}{2014 version} \\
\hline
 & Si-only & P-only & Si/P & Total \\
Number of structures & 244 & 65 & 74 & 383 \\
Number of frameworks & 131 & 43 & 30 & 204 \\
\hline
\multicolumn{5}{l}{2025 version} \\
\hline
 & Si-only & P-only & Si/P & Total \\
Number of structures & 343 & 99 & 81 & 523 \\
Number of frameworks & 163 & 53 & 30 & 246 \\
\hline
\end{tabular}
\end{table}

\subsection{Model and training}
For all modeling work, PyTorch v2.3.1 was used with an Nvidia RTX 2080TI or A100 GPU as the accelerator. All models were trained to predict the classification of IZA and PCOD structures using the cross entropy as the loss function. To deal with the unbalanced dataset, WeightedRandomSampler was used in DataLoader during training to sample from each class in proportion to the inverse of the number of samples in that class. The volumetric distance grids used as the input for ZeoNet is constructed using a probe with a radius of 1.2~\AA.\cite{liu2023zeonet,liu2025representation}

To compute the formation energy, the structures were relaxed using the General Utility Lattice Program (GULP)\cite{gale2003general} with the modified SLC (Sanders–Leslie–Catlow) force field.\cite{sanders1984m, helfrecht2022ranking} To visualize the ZeoNet-based classifiers, the t-distributed stochastic neighbor embedding (t-SNE) was applied to reduce the 512-dimensional learned embedding to a two-dimensional space.\cite{JMLR:v9:vandermaaten08a}

\subsection{Geometric filters}
The local interatomic distances (LIDs) were computed using Pymatgen,\cite{ong2013pymatgen} and used to construct 11 filter criterion based on the comparison with IZA zeolites.\cite{salcedo2019high} Some of these criteria require parameters that are determined by fitting to the distribution of IZA zeolites; see Figure~S4 and Table~S3 for the determination and tabulated values of $C_{1,1}$, $C_{1,2}$, $C_{2,1}$, $C_{2,2}$, $C_{3,1}$, and $C_{3,2}$. These 11 criteria are listed below, where $\sigma_{d_\mathrm{AB}}$ and $R_{d_\mathrm{AB}}$ denote the standard deviation and range of the corresponding LIDs $d_\mathrm{AB}$.

\begin{itemize}
    \item Category I
\[
\epsilon_{\mathrm{mean}\ d_\mathrm{OO}} = \left| \mathrm{mean}\ d_\mathrm{OO} - C_{1,1} \times \mathrm{mean}\ d_\mathrm{TO} - C_{1,2} \right| \leq \max\left( \epsilon_{\mathrm{mean}\ d_\mathrm{OO}} \right)_{\mathrm{IZA}}
\]
\[
\epsilon_{\mathrm{mean}\ d_\mathrm{TT}} = \left| \mathrm{mean}\ d_\mathrm{TT} - C_{2,1} \times \mathrm{mean}\ d_\mathrm{TO} - C_{2,2} \right| \leq \max\left( \epsilon_{\mathrm{mean}\ d_\mathrm{TT}} \right)_{\mathrm{IZA}}
\]
\[
\epsilon_{\mathrm{min}\ d_\mathrm{T5}} = \left| -\mathrm{min}\ d_\mathrm{T5} + C_{3,1} \times \mathrm{min}\ d_\mathrm{TT} + C_{3,2} \right| \leq \max\left( \epsilon_{\mathrm{min}\ d_\mathrm{T5}} \right)_{\mathrm{IZA}}
\]

    \item Category II
\[
\sigma_{d_\mathrm{TT}} \leq \max\left(\sigma_{d_\mathrm{TT}}\right)_{\mathrm{IZA}}
\]
\[
\sigma_{d_\mathrm{TO}} \leq \max\left(\sigma_{d_\mathrm{TO}}\right)_{\mathrm{IZA}}
\]
\[
\sigma_{d_\mathrm{OO}} \leq \max\left(\sigma_{d_\mathrm{OO}}\right)_{\mathrm{IZA}}
\]
\[
\sigma_{d_\mathrm{T5}} \leq \max\left(\sigma_{d_\mathrm{T5}}\right)_{\mathrm{IZA}}
\]

    \item Category III
\[
R_{d_\mathrm{TT}} \leq \max\left(R_{d_\mathrm{TT}}\right)_{\mathrm{IZA}}
\]
\[
R_{d_\mathrm{TO}} \leq \max\left(R_{d_\mathrm{TO}}\right)_{\mathrm{IZA}}
\]
\[
R_{d_\mathrm{OO}} \leq \max\left(R_{d_\mathrm{OO}}\right)_{\mathrm{IZA}}
\]
\[
R_{d_\mathrm{T5}} \leq \max\left(R_{d_\mathrm{T5}}\right)_{\mathrm{IZA}}
\]
\end{itemize}

\section{Supporting Information}
The Supporting Information is available free of charge at \\https://pubs.acs.org/doi/10.1021/acsmaterialslett.

Additional figures showing the lattice energy distribution, geometric analyses, t-SNE plot;

Tables listing misclassified IZA and PCOD structures and the classification of IZA zeolite framework topologies;

An interactive web application for exploring the dataset and model predictions: \\https://r.bai.group/zeolite-classifier.

\section{Acknowledgements}
This project was supported by the Defense Advanced Research Projects Agency (DARPA) under Grant No. D24AP00322-00. Funding for the GPU cluster used in this work was provided by the National Science Foundation under Grant No. CNS-1919334. The research also used resources of the Advanced Cyberinfrastructure Coordination Ecosystem: Services \& Support (ACCESS) through Allocation No. CHM250039.

\section{Conflict of Interest}
The authors declare no conflict of interest.

\section{Data Availability Statement}
The data that support the ﬁndings of this study are available from Supporting Information.


\bibliography{ref}
\end{document}


\newpage

\section{Additional Figures}
\begin{figure}[htb]
    \centering
    \includegraphics[width=0.8\textwidth]{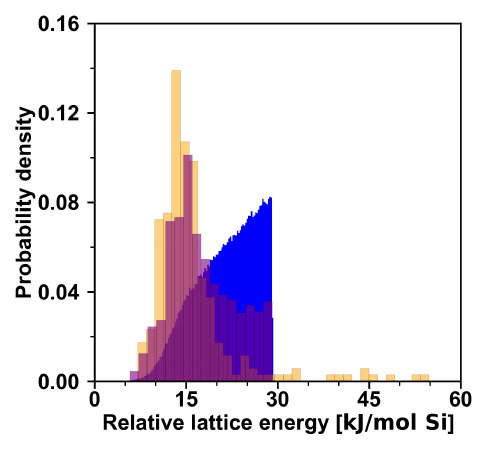}
    \caption{Probability density for the distributions of formation energy for IZA zeolites (orange), all PCOD structures (blue), and misclassified PCOD structures (purple). The energy of the most stable structure was set to be zero.}
    \label{fig:energy_distribution}
\end{figure}

\begin{figure}[htb]
    \centering
    \includegraphics[width=0.8\textwidth]{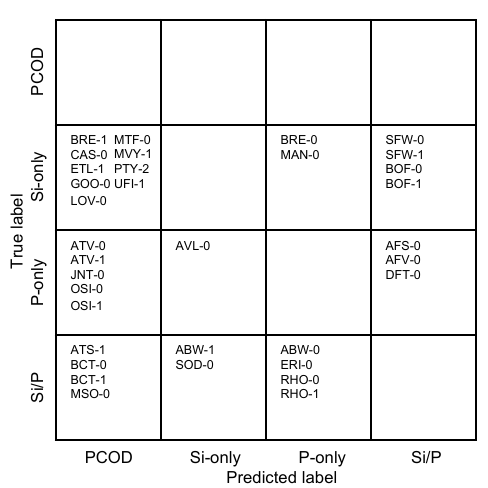}
    \caption{List of 18 IZA zeolites misclassified by the four-class model.}
    \label{fig:mis-iza}
\end{figure}

\begin{figure}[htb]
    \centering
    \includegraphics[]{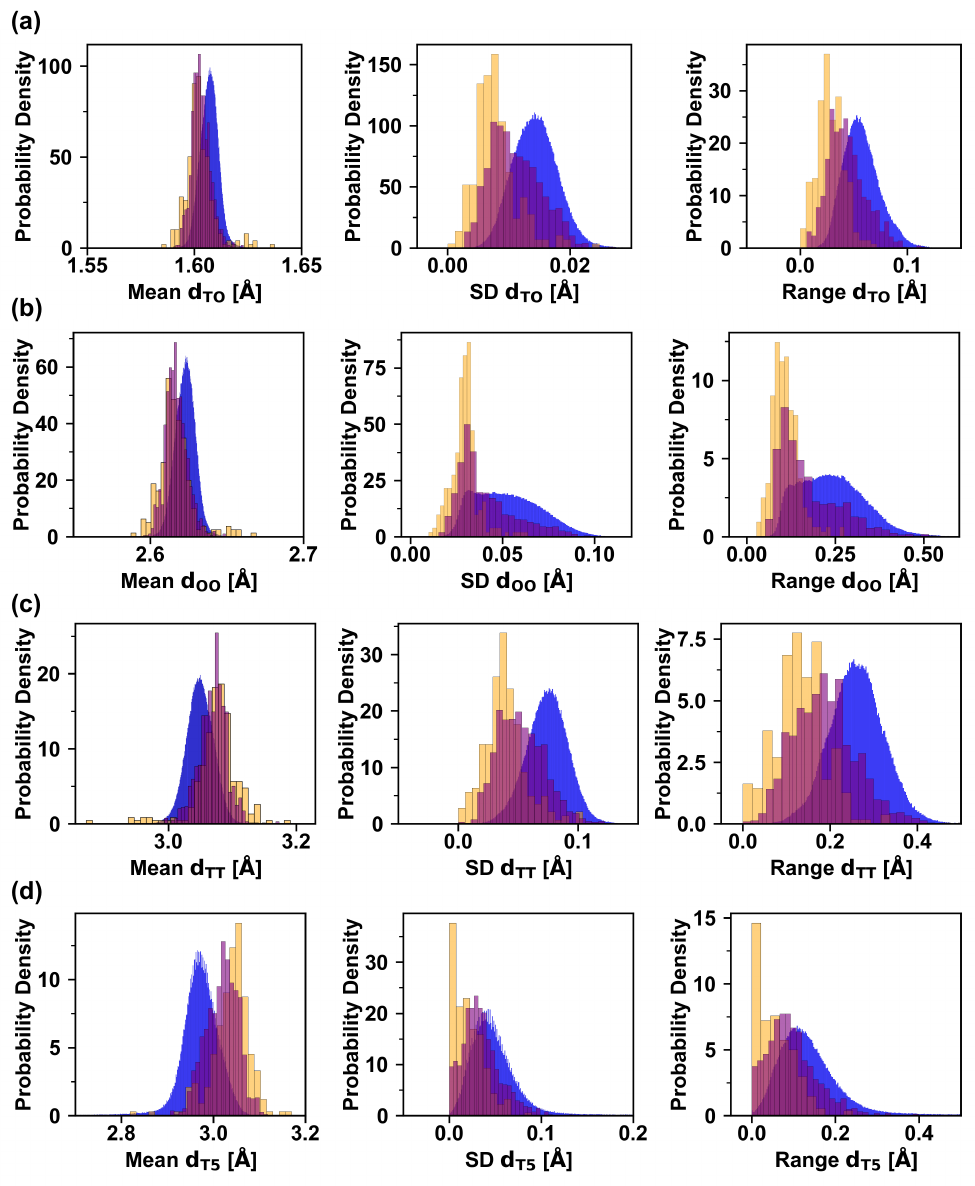}
    \caption{Probability density for the distributions of mean, standard deviation, and range of four local interatomic distances: (a) T–O distance, (b) O–(T)–O distance, (c) T–(O)–T distance and (d) distance to the fifth nearest neighbor of a T-atom for IZA zeolites (orange), all PCOD structures (blue), and misclassified PCOD structures (purple).}
    \label{fig:lids_distribution}
\end{figure}

\begin{figure}[htb]
    \centering
    \includegraphics[]{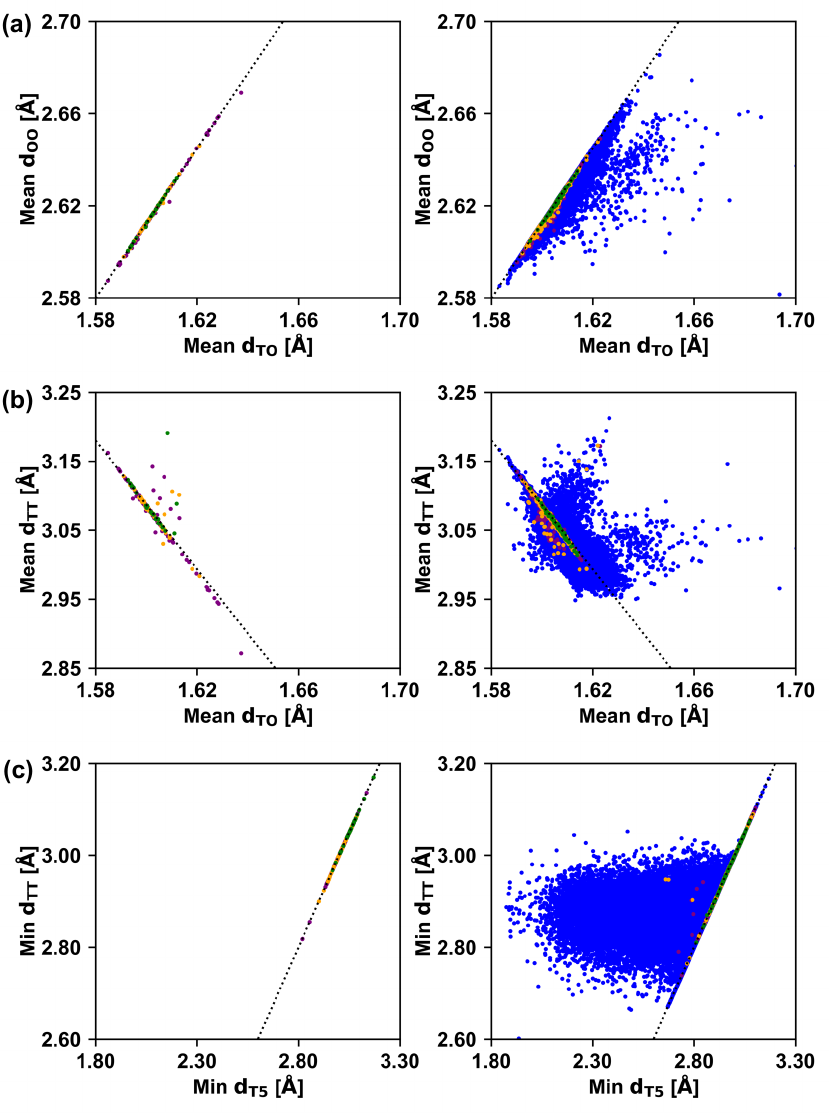}
    \caption{Linear fits for selected pairs of local interatomic distances: (a) mean O–(T)–O distance vs. mean T–O distance, (b) mean T–(O)–T distance vs. mean T–O distance, and (c) shortest T–(O)–T distance vs. shortest distance to the fifth nearest neighbor of a T-atom. For each plot, the left panel shows IZA zeolites and their linear fit (dashed lines) while the right panel shows all PCOD structures (blue) and the same IZA-based linear fit (dashed lines). IZA and misclassified PCOD structures are colored in purple, orange, and green, respectively, for Si-only, P-only, and Si/P structures.}
    \label{fig:fit}
\end{figure}

\begin{figure}[htb]
    \centering
    \includegraphics[width=0.8\textwidth]{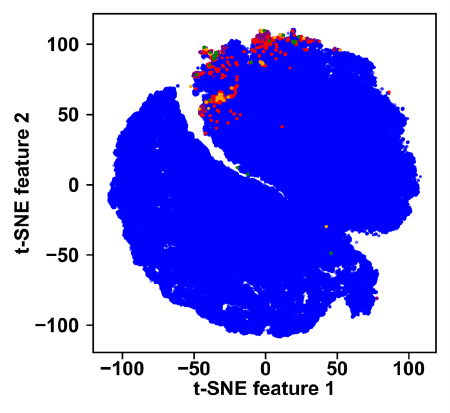}
    \caption{Two-dimensional t-SNE embedding of the four-class model. The ZeoNet embedding is a 512-dimensional vector. IZA structures are colored in purple, orange, and green, respectively, for Si-only, P-only, and Si/P structures. PCOD and misclassified PCOD structures are shown in blue and red, respectively.}
    \label{fig:tsne}
\end{figure}

\begin{figure}[htb]
    \centering
    \includegraphics[]{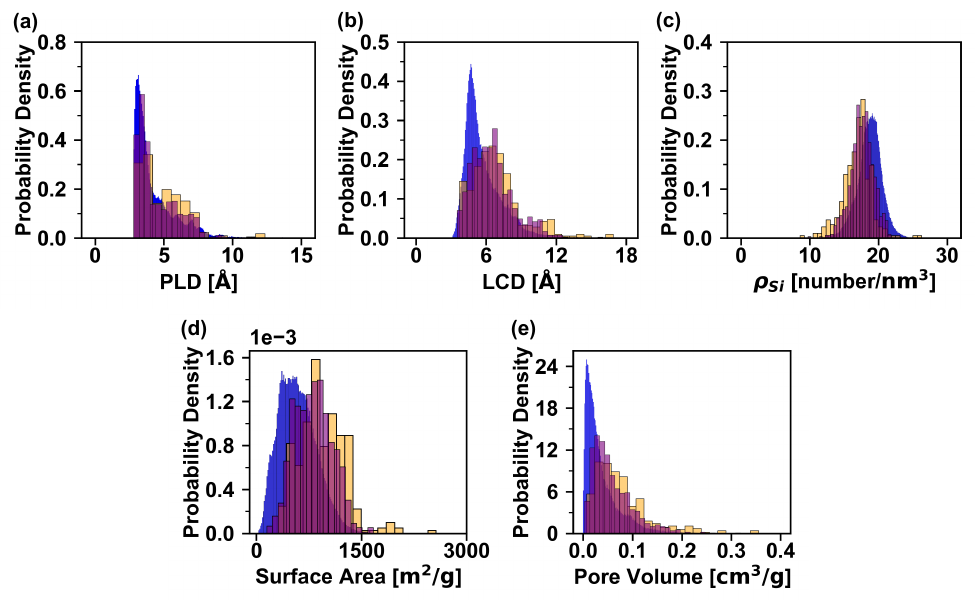}
    \caption{Probability density for the distributions of five geometric features: (a) pore-limiting diameter (PLD), (b) largest-cavity diameter (LCD), (c) T atom number density ($\rho_\textrm{Si}$), (d) surface area, and (e) pore volume for IZA zeolites (orange), all PCOD structures (blue), and misclassified PCOD structures (purple).}
    \label{fig:geometry}
\end{figure}

\clearpage

\section{Additional Tables}

\begin{table}[]
\caption{Classification of IZA zeolite frameworks by their primary tetrahedral atoms. Each IZA framework code is assigned to one of three categories: Si-only, P-only, or Si/P. '*' indicates the disordered framework structures with intergrowths.}
\centering
\label{tab:class_count}
\begin{tabular}{ccccccccccc}
\multicolumn{10}{c}{Si-only (162)} \\
AFG     & ANA & ASV & *BEA & BEC & BIK & BOF & BOG & BOZ & BRE  \\
BSV     & CAS & CDO & CFI & CON & CSV & DAC & DDR & DOH & DON \\
EAB     & EEI & EMT & EON & EOS & EPI & ESV & ETL & ETR & ETV \\
EUO     & EWF & EWO & EWS & FAR & FER & FRA & GIU & GON & GOO \\
HEU     & HZF & IFR & IFW & IFY & IHW & IMF & IRN & IRR & ISV \\
ITE     & ITG & ITH & ITR & ITT & ITW & IWR & IWS & IWV & IWW \\
JBW     & JOZ & JSR & JST & JZO & JZT & KFI & LIO & LOV & LTF \\
LTJ     & LTL & LTN & MAN & MAR & MAZ & MEL & MEP & MFI & MFS \\
MON     & MOR & MOZ & *MRE & MRT & MSE & MTF & MTN & MTT & MTW \\
MVY     & MWF & MWW & NAB & NAT & NES & NON & NPO & NPT & NSI \\
OBW     & OFF & OKO & OSO & PAU & PCR & POS & PTF & PTO & PTT \\
PTY     & PUN & PWN & PWO & PWW & RFE & RRO & RSN & RTE & RTH \\
RUT     & RWR & RWY & SBN & SEW & SFE & SFF & SFG & SFH & SFN \\
SFS     & *SFV & SFW & SGT & SOF & SOR & SOS & SOV & SSF & SSY \\
STF     & STI & *STO & STT & STW & SVV & SZR & TER & TOL & TON \\
TSC     & TUN & UFI & UOS & UOV & UOZ & UTL & UWY & VET & VNI \\
VSV     & YFI & YUG &     &     &     &     &     &     &     \\
\multicolumn{10}{c}{P-only (53)} \\
ACO     & AEL & AEN & AET & AFN & AFO & AFR & AFS & AFV & AFY \\
AHT     & ANO & APC & APD & ATN & ATO & ATV & AVE & AVL & AWO \\
AWW     & CGF & CZP & DFO & DFT & EZT & IFO & JNT & JRY & JSN \\
JSW     & JSY & OSI & OWE & PON & POR & PSI & SAF & SAO & SAT \\
SAV     & SBE & SBS & SBT & SFO & SIV & SWY & UEI & USI & VFI \\
WEI     & ZJN & ZON &     &     &     &     &     &     &     \\
\multicolumn{10}{c}{Si/P (31)} \\
ABW     & AEI & AFI & AFT & AFX & AST & ATS & ATT & BCT & BPH\\
CAN     & CGS & CHA & EDI & ERI & FAU & GIS & GME & LAU & LEV \\
LOS & LTA & MEI & MER & MSO & PHI & RHO & SAS & SOD & THO \\
\end{tabular}
\end{table}

\newpage
\begin{longtable}{lllll|lllll}
\caption{List of PCOD structures misclassified by the four-class model and their normalized softmax probabilities for each class.} 
\label {tab:4class_candidates} \\
\hline
Zeolite & $P_{\mathrm{PCOD}}$ & $P_{\mathrm{Si-only}}$ & $P_{\mathrm{P-only}}$ & $P_{\mathrm{Si/P}}$ & Zeolite & $P_{\mathrm{PCOD}}$ & $P_{\mathrm{Si-only}}$ & $P_{\mathrm{P-only}}$ & $P_{\mathrm{Si/P}}$ \\
\hline
\endfirsthead

\hline
Zeolite & $P_{\mathrm{PCOD}}$ & $P_{\mathrm{Si-only}}$ & $P_{\mathrm{P-only}}$ & $P_{\mathrm{Si/P}}$ & Zeolite & $P_{\mathrm{PCOD}}$ & $P_{\mathrm{Si-only}}$ & $P_{\mathrm{P-only}}$ & $P_{\mathrm{Si/P}}$ \\
\hline
\endhead

\hline
\endfoot
8114334 & 0    & 0    & 1    & 0    & 8297975 & 0.09 & 0    & 0.91 & 0    \\
8325316 & 0    & 1    & 0    & 0    & 8177960 & 0.09 & 0    & 0.91 & 0    \\
8322698 & 0    & 0    & 1    & 0    & 8304316 & 0.09 & 0.91 & 0    & 0    \\
8285304 & 0    & 1    & 0    & 0    & 8264673 & 0.09 & 0.91 & 0    & 0    \\
8234944 & 0    & 0    & 1    & 0    & 8123219 & 0.1  & 0.9  & 0    & 0    \\
8056793 & 0    & 0    & 0    & 1    & 8030566 & 0.1  & 0.9  & 0    & 0    \\
8114396 & 0    & 1    & 0    & 0    & 8183149 & 0.1  & 0    & 0.9  & 0    \\
8065514 & 0    & 1    & 0    & 0    & 8142998 & 0.1  & 0.9  & 0    & 0    \\
\end{longtable}

\begin{table}[htb]
    \centering
 \begin{minipage}[t]{0.45\linewidth}
    \vspace{0pt}
    \begin{tabular*}{\textwidth}{@{\extracolsep{\fill}}cc}
      \toprule
    Parameter & Value \\
    \midrule
    $C_{1, 1}$ & 1.623 \\
    $C_{1, 2}$ & $1.518 \times 10^{-2}$ \\
    $C_{2, 1}$ & $-4.659$ \\
    $C_{2, 2}$ & 10.54 \\
    $C_{3, 1}$ & 1.000 \\
    $C_{3, 2}$ & $1.359 \times 10^{-15}$ \\ 
    \midrule
    max($\mathrm{\epsilon_{mean\ d_{OO}}}$) & $5.377 \times 10^{-3}$  ~\AA\ \\
    max($\mathrm{\epsilon_{mean\ d_{TT}}}$) & $1.428 \times 10^{-1}$ ~\AA\ \\
    max($\mathrm{\epsilon_{min\ d_{T5}}}$) & $1.305 \times 10^{-15}$ ~\AA\ \\
      \bottomrule
    \end{tabular*}
  \end{minipage}
  \hspace{1em}
  \begin{minipage}[t]{0.45\linewidth}
    \vspace{0pt}
    \begin{tabular*}{\textwidth}{@{\extracolsep{\fill}}cc}
      \toprule
    Parameter & Value \\
    \midrule
    max($\mathrm{\sigma_{d_{TT}}}$) & $1.038 \times 10^{-1}$ ~\AA\ \\ 
    max($\mathrm{\sigma_{d_{TO}}}$) & $2.481 \times 10^{-2}$ ~\AA\ \\
    max($\mathrm{\sigma_{d_{OO}}}$) & $6.382 \times 10^{-2}$ ~\AA\ \\
    max($\mathrm{\sigma_{d_{T5}}}$) & $8.940 \times 10^{-2}$ ~\AA\ \\
    \midrule
    max($\mathrm{R_{d_{TT}}}$) & $3.382 \times 10^{-1}$ ~\AA\ \\
    max($\mathrm{R_{d_{TO}}}$) & $7.158 \times 10^{-2}$ ~\AA\ \\
    max($\mathrm{R_{d_{OO}}}$) & $2.756 \times 10^{-1}$ ~\AA\ \\
    max($\mathrm{R_{d_{T5}}}$) & $2.416 \times 10^{-1}$ ~\AA\ \\  
      \bottomrule
    \end{tabular*}
  \end{minipage}
  \caption{Threshold and fitted parameters used in the geometric filters}
  \label{fig:citeria}
\end{table}

\clearpage